\documentclass[twocolumn,superscriptaddress,nofootinbib, floatfix, amsfonts, aps]{revtex4-1}
\usepackage{amsmath}
\usepackage{bm}
\usepackage{graphicx}
\usepackage{subfigure}
\usepackage{float}
\usepackage{mathrsfs}
\usepackage{footmisc}
\usepackage{multirow}
\usepackage{graphicx}
\usepackage{soul,xcolor}
\usepackage{xcolor, ulem, color, framed}
\usepackage{amssymb}
\usepackage{setspace}
\usepackage[colorlinks, linkcolor=red,anchorcolor=green,citecolor=blue]{hyperref}

\graphicspath{{./figs/}}

\begin{document}

\title{Bottomonium evolution with in-medium heavy quark potential from lattice QCD}

\author{Ge Chen}
\affiliation{Department of physics and Astronomy, UCL, Gower St, London WC1E 6BT, England}
\author{Baoyi Chen}
\affiliation{Department of Physics, Tianjin University, Tianjin 300354, China}
\author{Jiaxing Zhao}
\affiliation{SUBATECH, Universit\'e de Nantes, IMT Atlantique, IN2P3/CNRS, 4 rue Alfred Kastler, 44307 Nantes cedex 3, France}

\date{\today}

\begin{abstract}
The static properties and dynamic evolution of bottomonium states in a hot QCD medium are investigated through the Schr\"odinger equation with complex heavy quark potentials, which are presented recently in lattice QCD study and with three different extractions. 
This approach builds a direct connection between the in-medium heavy quark potentials from the lattice QCD to the experimental observables. 
The yields and nuclear modification factors $R_{AA}$ of bottomonium in Pb-Pb collisions at $\sqrt{s_{\rm NN}}=5.02~\rm TeV$ are calculated in this work.
Our results show a large suppression of the bottomonium yield in heavy ion collisions due to the large imaginary potential. To understand bottomonium $R_{AA}$ based on lattice QCD potentials, we propose a fomration time for bottomonium states and find that experimental data can be well explained with the heavy quark potential extracted by the Pad\'e fit, which shows no color screening in the real part potential. 
\end{abstract}
\maketitle

It is widely believed that relativistic heavy-ion collisions, performed at the Relativistic Heavy Ion Collider (RHIC) and Large Hadron Collider (LHC), can generate the deconfined Quantum Chromodynamic (QCD) matter called Quark-Gluon Plasma (QGP)~\cite{Bazavov:2011nk,Busza:2018rrf}. Extensive research has been conducted to study the properties of the QGP at different temperatures and baryon chemical potentials.
Quarkonia, the bound state of heavy quark and its antiquark, is an excellent probe to study the QGP properties. Besides, quarkonia can be used to examine the QCD at finite temperatures. Due to the hierarchy of energy scales $m_Q\gg m_Qv \gg m_Qv^2$ of heavy quarks in vacuum, the quarkonia can be described by, for example, the non-relativistic QCD (NRQCD) and potential non-relativistic QCD (pNRQCD) theories~\cite{Caswell:1985ui,Brambilla:1999xf,Brambilla:2004jw}. The equation of motion in pNRQCD returns to a Schr\"odinger-like equation with the transition between color singlet state and octet state~\cite{Brambilla:2004jw}, where the potential at the leading order for singlet state reads $V(r)=-C_F\alpha_s/r$ with $C_F=(N_c^2-1)/(2N_c)$ and the number of colors $N_c=3$. Furthermore, the non-perturbative lattice simulations show a linear potential between a heavy quark-antiquark pair at long range, and the full range potential is very close to the Cornell potential~\cite{Kawanai:2011jt}. Employing the two-body Schr\"odinger equation with the Cornell potential, the quarkonium mass spectra can be described successfully, see the review paper~\cite{Zhao:2020jqu}.

In the hot medium, the perturbative investigation of quarkonium properties depends on the relationship of new scales, e.g. temperature $T$, Debye screening mass $m_D$. The previous perturbative studies indicate that in most cases, quarkonium properties in the hot medium are determined by a finite-temperature potential, which has an imaginary part~\cite{Brambilla:2008cx,Brambilla:2010vq}. The real part of the potential reveals the color screening effect in the hot medium~\cite{Karsch:1987pv}, while the imaginary part is introduced by the Landau damping or color singlet to octet transition~\cite{Brambilla:2008cx}.
At extremely high temperatures, the Landau damping becomes dominant, and the potential can be obtained by the Hard-Thermal Loop (HTL) resummed perturbation~\cite{Laine:2006ns,Beraudo:2007ky}. 
In the strong coupling regime, various methods such as lattice QCD~\cite{Rothkopf:2011db,Lafferty:2019jpr,Bazavov:2018wmo,Petreczky:2017aiz,Bala:2021fkm} and the T-matrix method~\cite{Liu:2017qah} have been employed to investigate the heavy quark potential. 

With quarkonium thermal properties, their production in heavy ion collisions has been approached through various methods. The classical transport approach~\cite{Zhao:2010nk,He:2021zej,Du:2022uvj,Yan:2006ve,Liu:2017qah,Zhou:2014kka,Chen:2018kfo,Yao:2018nmy,Yao:2018sgn}, quantum method such as time-dependent Schr\"odinger equation~\cite{Katz:2015qja,Islam:2020bnp,Islam:2020gdv,Wen:2022yjx,Wen:2022utn}, the improved Remler equation~\cite{Villar:2022sbv,Song:2023zma}, and open quantum system approach~\cite{Borghini:2011ms,Blaizot:2015hya,Blaizot:2017ypk,Blaizot:2018oev,Akamatsu:2011se,Akamatsu:2012vt,Kajimoto:2017rel,Delorme:2022hoo,Brambilla:2016wgg,Brambilla:2023hkw}. The final yield and transverse momentum spectrum reveal the in-medium binding energy and thermal width of quarkonium, which are encoded in the real and imaginary part of the potential. In the framework of the time-dependent Schr\"odinger equation, the color screening effect encoded in the real part of the potential results in transitions between different bound states or scattering states, while the imaginary part of the potential suppresses the wave function of singlet states which can be attributed to the transition from singlet to octet states. As components of octet states are usually not included in the Schr\"odinger equation, those singlet-octet state transitions reduce the normalization of the wave function. In previous studies, the real part of the in-medium potential is suggested to locate between the free energy and the internal energy, both of which show a clear color screening effect as temperature increases, while the imaginary potential is given by the HTL or quenched lattice QCD~\cite{Islam:2020bnp,Islam:2020gdv,Wen:2022yjx,Wen:2022utn}. However, recent lattice QCD calculations with dynamic quarks suggest that there are no color screening effects for the real part heavy quark potential, even at high temperatures, e.g. $T\le 3T_c$ with the critical temperature $T_c$ and $r \le 1~ \rm fm/c$~\cite{Bazavov:2023dci,Bala:2021fkm}, which is larger than bottomonium averaged radius~\cite{Zhao:2020jqu}. In the meantime, the imaginary part is much larger than that from the HTL and quenched lattice QCD. This ``novel'' heavy quark potential is also supported by other studies, such as the heavy potential obtained with the Gribov-Zwanziger approach~\cite{Wu:2022nbv,Debnath:2023dhs}, machine learning method~\cite{Shi:2021qri}.   
In this paper, we utilize the heavy quark finite-temperature potential, which has been obtained recently in lattice QCD, to study the bottomonium static properties and dynamical evolution in heavy ion collisions. 

\begin{figure*}[!htb]
\includegraphics[width=0.85\textwidth]{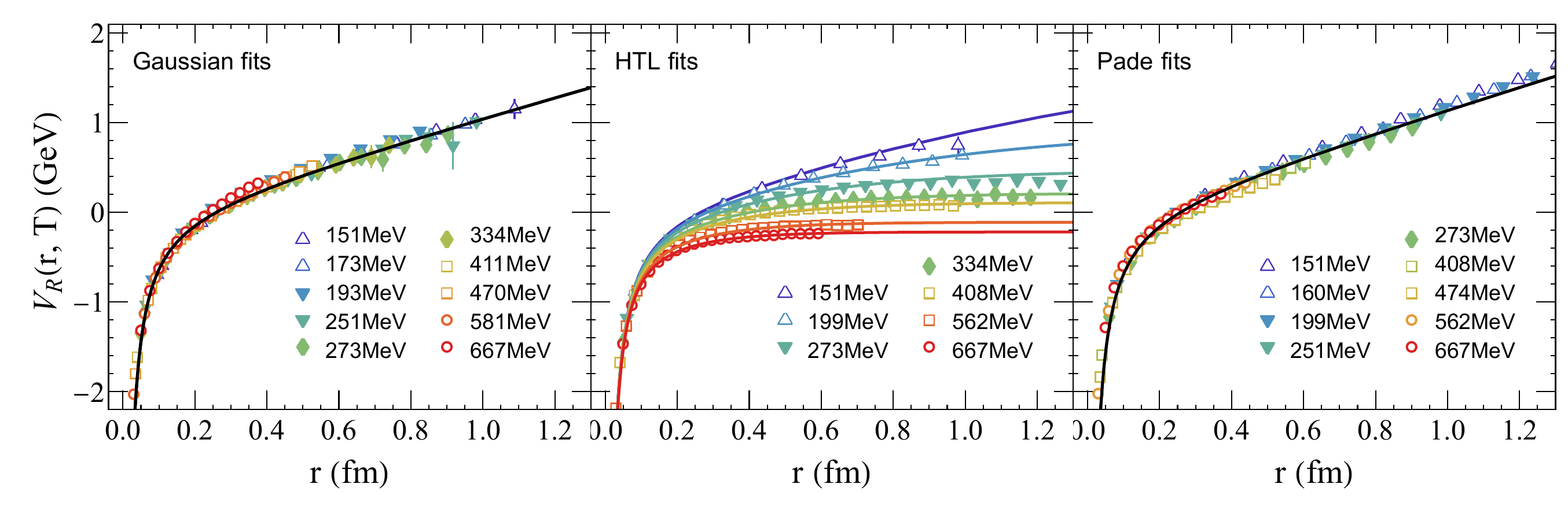}\\
\includegraphics[width=0.85\textwidth]{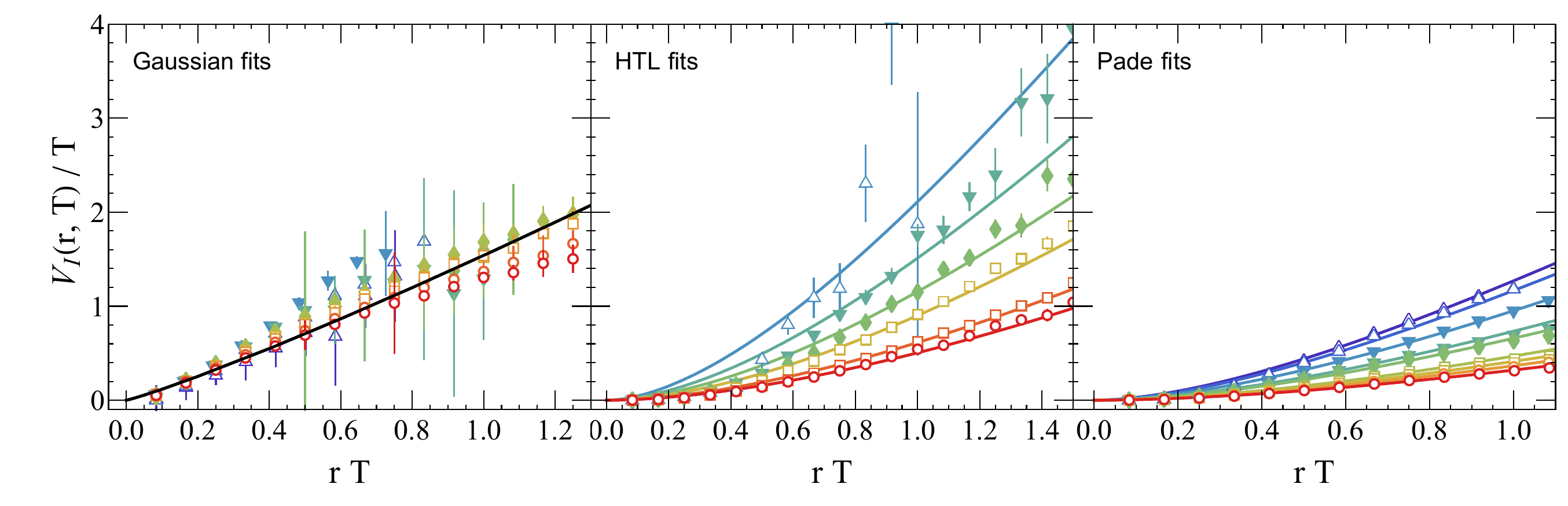}
\caption{The real and imaginary part potential between heavy quark $Q$ and $\bar Q$ at finite temperature. From left to right are the potential obtained from the Gaussian fits, the HTL fits, and the Pad\'e fits. The data are from the lattice QCD simulation~\cite{Bala:2021fkm}. The curves are the fitting results with Eq.~\eqref{eq.gauss},~\eqref{eq.htl}, and~\eqref{eq.pade}.}
\label{fig.ReImV}
\end{figure*}

In lattice QCD, the heavy quark potential can be extracted from the spectrum function. In order to constrain the spectral function from limited data on Euclidean time correlation functions, one needs to assume some functional form for it. The spectral function has a low frequency tail, a temperature-independent high frequency term, and a peak structure at intermediate frequency regions.   
Recent studies employed several different functional forms to extract the spectrum function, such as the Gaussian fit (a Gaussian shape for the peak), the HTL fit (the peak position and peak width are considered as the real and imaginary part of thermal static energy, which is given by HTL method), and Pad\'e fit (a model independent approach and based on the Pad\'e rational approximation). The results
indicate that the extraction method significantly influences the resulting heavy quark potential~\cite{Bala:2021fkm}.
In the following, we will introduce these three different potentials obtained by above mentioned extraction methods. The first one is extracted by the Gaussian fits method. The complex potential can be separated as a real part and an imaginary part, $V(r, T)=V_R(r,T)-iV_I(r,T)$. The real and imaginary parts of the potential are shown in the left of Fig.~\ref{fig.ReImV}. The real part potential shows non-screening and the temperature-reduced imaginary part potential has a same trend, which can be parameterized as,
\begin{eqnarray}
V_R^{\rm Gauss}(r,T) &=&-{\alpha_1 \over r} + \sigma_1 r, \nonumber\\
V_I^{\rm Gauss}(r,T) &=&T[{(rT)}^{\beta_1} + \gamma_1 {(rT)}],
\label{eq.gauss}
\end{eqnarray}
where the parameters $\alpha_1=0.3805$ and $\sigma_1=0.22\rm GeV^2$. $\beta_1=1.2$ and $\gamma_1=0.54$.  
Besides, the real and imaginary potential obtained via the HTL extraction~\cite{Bala:2021fkm} show huge difference where an obvious screening effect appears. The lattice data and the fitting curves are shown in the middle of Fig.~\ref{fig.ReImV}. The parameterization form can be expressed as,
\begin{eqnarray}
V_R^{\rm HTL}(r,T) & =& -\alpha_2 \left( m_{\rm DR}^{\rm HTL} + {e^{-m_{\rm DR}^{\rm HTL} r} \over r}\right) \nonumber\\
&+& {2\sigma_2\over m_{\rm DR}^{\rm HTL}} -  {e^{-m_{\rm DR}^{\rm HTL}r} (2 + m_{\rm DR}^{\rm HTL} r)\sigma_2\over m_{\rm DR}^{\rm HTL}}, \nonumber\\
V_I^{\rm HTL}(r,T) & =&\gamma_2\alpha_2 T \phi(m_{\rm DI}^{\rm HTL} r),
\label{eq.htl}
\end{eqnarray}
where $\alpha_2=0.3805$, $\sigma_2=0.22\rm GeV^2$, $\gamma_2=100$. The Debye screening mass in real ($m_{\rm DR}^{\rm HTL}$) and imaginary part ($m_{\rm DI}^{\rm HTL}$) can be obtained by fitting the lattice data. 
The definition of $\phi$ is 
\begin{eqnarray}
\phi(x)=2\int_0^\infty dz{z\over (z^2+1)^2}\left( 1-{\sin(xz)\over xz} \right).
\end{eqnarray}

The third one is obtained via the Pad\'e fit as shown in the right of Fig.~\ref{fig.ReImV}, which also has a non-screening real part potential. The parameterization of $V(r,T)$ can be expressed as,
\begin{eqnarray}
V_R^{\rm Pade}(r,T) &=& - {\alpha_3\over r} + \sigma_3 r, \nonumber\\
V_I^{\rm Pade}(r,T) & =&\gamma_3 \alpha_3 T \phi(m_{\rm DI}^{\rm Pade} r),
\label{eq.pade}
\end{eqnarray}
where $\alpha_3=0.41$, $\sigma_3= 0.24\rm GeV^2$, and $\gamma_3=60$. The Debye screening mass in the imaginary part $m_{\rm DI}^{\rm Pade}$ can be got from the lattice data. 

\begin{figure}[!htb]
\includegraphics[width=0.4\textwidth]{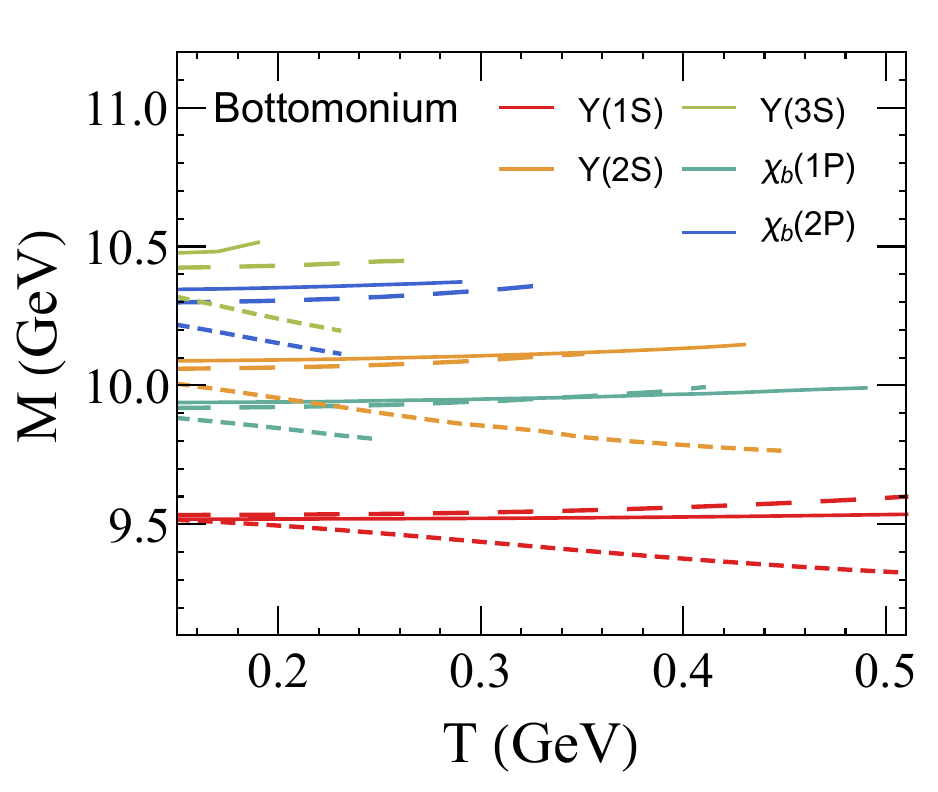}\\
\includegraphics[width=0.4\textwidth]{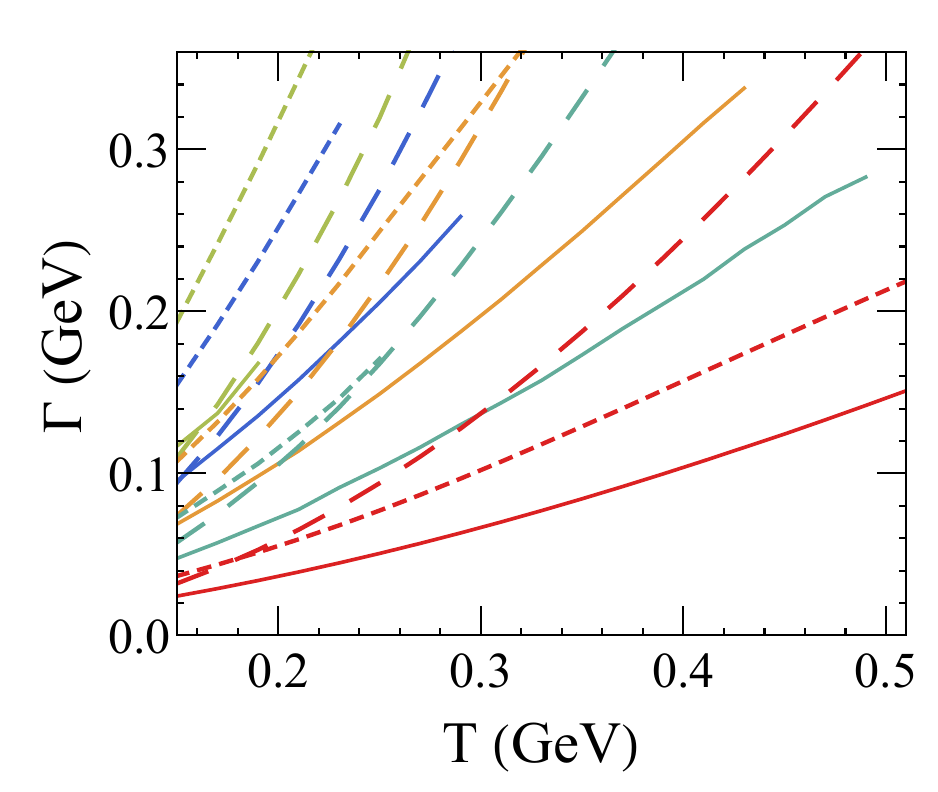}\\
\includegraphics[width=0.4\textwidth]{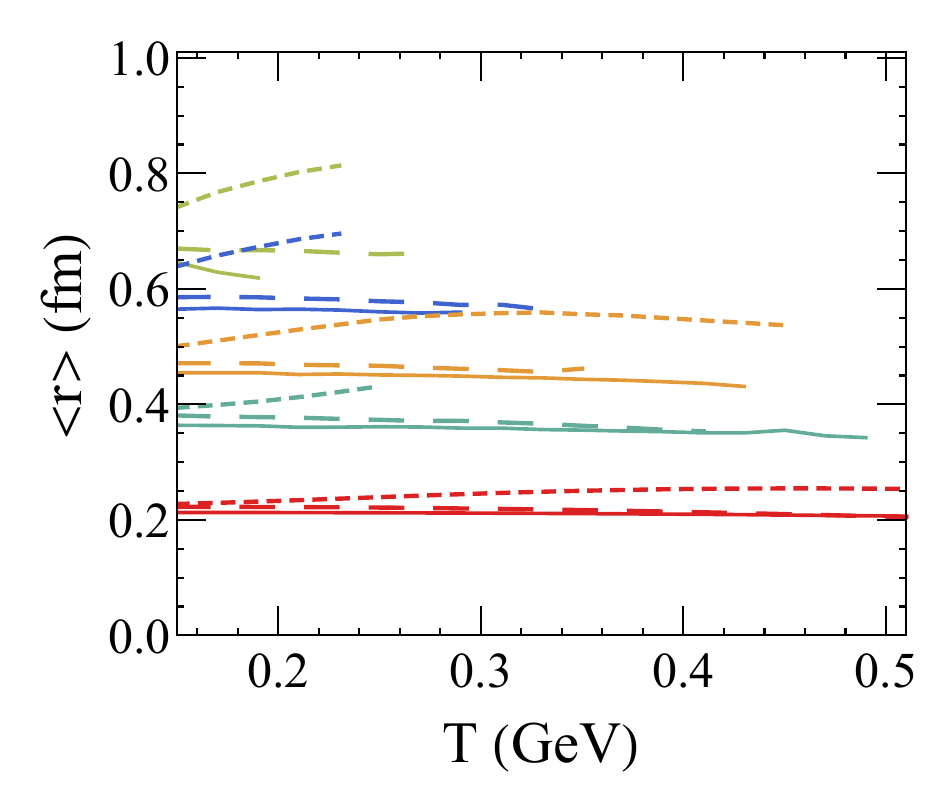}
\caption{
The masses, widths, and radii of various bottomonium states under different finite-temperature potentials are depicted in Fig.~\ref{fig.ReImV}. The potential obtained from the Gaussian fits is represented by the long-dashed lines, the HTL fits by the short-dashed lines, and the Pad\'e fits by the solid lines.
}
\label{fig.mass_width}
\end{figure}
The static properties of the bottomonium states in a hot medium can be calculated by the Schr\"odinger equation with the above-mentioned finite-temperature potential at given temperature. Because the potential is angle-independent, the center-of-mass motion can be separated out and the reduced two-body Schr\"odinger equation reads,
\begin{eqnarray}
\left[-{\nabla^2 \over 2m_\mu}+V_R(r,T)-iV_I(r,T)\right]\Psi({\bm r})=E_r\Psi({\bm r}),
\end{eqnarray}
where $m_\mu=(m_bm_{\bar b})/(m_b +m_{\bar b})=m_b/2$ is the reduced mass and bottom quark mass $m_b=4.7~\rm GeV$, $\psi$ is relative wavefuntion of bottomonium states and $E_r$ is the eigen energies.
With the introduction of this complex potential, both the energy eigenvalues and wavefunctions acquire complex values. The real part of the energy eigenvalue determines the mass of the bottomonium state: $M(T)=2m_b+{\rm Re}[E_r(T)]$, while the imaginary part corresponds to the thermal width $\Gamma(T)=-{\rm Im}[E_r(T)]$. The averaged radius of the bottomonium state is defined as $\langle r\rangle=\int \Psi \Psi^* |r|d^3{\bm r}$. Figure~\ref{fig.mass_width} illustrates the masses, thermal widths, and radii of the bottomonium states up to the $3S$ state. It is observed that the potentials obtained from the Gaussian fits and Pad\'e fits, which lack the screening effect, have almost no impact on the real part of the potential. Consequently, the masses using these two potential models remain nearly constant and equivalent to the vacuum values. Conversely, the HTL fits yield a potential with strong screening, resulting in a decrease in mass until the bound state disappears and an increase in the averaged radius. As for the thermal widths, there are noticeable differences among the three potential models. The larger imaginary potential derived from the Gaussian fits gives rise to a wider thermal width compared to the other two models.

With these static properties in the hot medium, we now come to study the dynamic evolution and production of bottomonium states in relativistic heavy ion collisions.
The quark gluon plasma has been observed to be a strongly-coupled medium created in relativistic heavy ion collisions. Its dynamical expansion can be effectively simulated using hydrodynamic equations in conjunction with an appropriate equation of state (EoS). In this study, we utilized the MUSIC package~\cite{Schenke:2010rr,Schenke:2010nt} to simulate the expansion of hot medium and adopted the parametrization of the lattice QCD EoS with a smooth crossover between the QGP and the hadron resonance gas around critical temperature $T_c=170 ~ \rm MeV$ ~\cite{HotQCD:2014kol,Bernhard:2016tnd}. The initial temperature profiles of the hot medium were determined by analyzing the final multiplicity of charged hadrons. For the most central collisions and the central rapidity region, the maximum initial temperature of the medium was calculated to be $T_0(\tau_0=0.6,{\bf x}_T=0) = 510$ MeV at the center of the medium in $\sqrt{s_{\rm NN}}=5.02$ TeV Pb-Pb collisions~\cite{Zhao:2017yhj}. Here, $\tau_0=0.6$ fm/c refers to the start time of the hydrodynamics. An effective shear viscosity $\eta/s = 0.08$ and a zero bulk viscosity are chosen in this study~\cite{Bernhard:2016tnd}.

The crucial point is the formation time $\tau_\Upsilon$, also called the decoherence time of the bottomonium states, which is a quantum effect and still not clear. Different bottomonium states may have different formation times that can be estimated via $\tau_\Upsilon\propto 1/E_B$ with the binding energy of bottomonium state. So, a longer formation time is expected for higher excited states. There are no bottomonium states before $\tau_\Upsilon$ instead of a wave packet which is a superposition state of all bottomonium states. 
In previous transport approaches, either no dissociation or a reduced dissociation rate is taken when $\tau<\tau_\Upsilon$. 

Because the mass of bottom quarks is very large, the relative motion of bottom quarks in the bottomonium can be treated in a  nonrelativistic way. One can therefore employ the time-dependent Schr\"odinger equation to describe the evolutions of bottomonium wave functions with in-medium potentials after $\tau_\Upsilon$~\cite{Krouppa:2015yoa,Brambilla:2020qwo,Islam:2020bnp,Wen:2022utn}. The in-medium potential is almost isotropic when neglecting the shear-viscosity correction. So, the center-of-mass momentum is conserved and the center-of-mass motion can be separated out.
Furthermore, the wavefunciton of the relative motion can be factorized into a radial part and angular part, $\Psi({\bm r},t)=R_{nl}(r,t)Y_{lm}(\theta, \phi)$.
The radial wavefunction of state $(n,l)$ satisfies the Schr\"odinger equation, 
\begin{eqnarray}
\label{fun-rad-sch}
&&i {\partial \over \partial t}\psi_{nl}(r, t)=\hat H \psi_{nl}(r,t),\\
&&\hat H=-{1\over 2m_\mu}{\partial ^2\over \partial r^2} +V_R(r,T)-iV_I(r,T) + {L(L+1)\over 2 m_\mu r^2},\nonumber
\end{eqnarray}
where $\psi_{nl}(r,t)$ is defined as $\psi_{nl}(r,t)=rR_{nl}(r,t)$. $L$ is the quantum number of the angluar momentum. $m_\mu$ is the reduced mass.
The in-medium potential can be described by the three aforementioned parametrizations. As a result of the medium-modified potentials, there are internal evolutions within the $b\bar b$ wave package, leading to transitions between different bottomonium eigenstates. To study these evolutions, we employ the Crank-Nicolson method to numerically solve the time-dependent Schr\"odinger equation.

Because the total momentum is conserved, $b\bar b$ wave package perform a uniform linear motion with the given initial momentum in the hot medium. In the heavy ion collisions, there are more than one $b\bar b$ package primordially produced at different positions with different total momentum. $b\bar b$ packages in nuclear collisions can be treated as a superposition of the distribution in proton-proton (p-p) collisions~\cite{Wen:2022yjx}.
The initial spatial distribution of the $b\bar b$ wave package is proportional to the density of binary collisions. The momentum distribution of the total momentum can be estimated by the observed momentum spectrum of $\Upsilon$ in p-p collisions and can be fitted via the formula,
\begin{eqnarray}
\label{eq:pp-input}
f_{pp}^{b\bar b}({\bm p}_T)
 &=& {d\bar N_{pp}^{b\bar b}\over 2\pi p_T dp_T}\nonumber\\
&=&{(n-1)\over \pi (n-2) \langle p_T^2\rangle_{pp}} \left[1+{p_T^2\over (n-2) \langle p_T^2
\rangle_{pp}}\right] ^{-n} 
\end{eqnarray}
with $n=2.5$ and the 
mean transverse momentum square at the central rapidity
is fitted to be 
$\langle p_T^2\rangle_{pp}=80\ \mathrm{(GeV/c)^2}$ at 5.02 TeV~\cite{Wen:2022yjx} based on the 
measurements of $\Upsilon(1S)$ in p-p collisions~\cite{LHCb:2014dei,LHCb:2012aa}.

When the $b\bar b$ wave package moves out of the QGP medium, the final fraction of bottomonium eigenstates labeled with the radial and angular quantum number $(n,l)$ is calculated by projecting the wave package to the corresponding vacuum eigenstate wave function $|c_{nl}|^2=|\langle \psi(r,t)|\psi_{nl}^{\rm vacuum}(r)\rangle|^2$.
The yield of bottomonium eigenstate $(n,l)$ in the nuclear collisions can be obtained by averaging over the spatial and momentum space~\cite{Wen:2022yjx}.
\begin{eqnarray}
\label{eq-dRAA}
&& \langle |c_{nl}(t)|^2\rangle = \\ 
&&{\int d{\bm x}_{T}d{\bm p}_{T} |c_{nl}(t, {\bm x}_{T}, 
{\bm p}_{T})|^2 f_{pp}^{b\bar b}({\bm p}_{T},{\bm b})n_{\rm coll}({\bm x}_T)\mathcal{R}({\bm x}_T)\over \int d{\bm x}_{T}d{\bm p}_{T} f_{pp}^{b\bar b}({\bm p}_{T})n_{\rm coll}({\bm x}_T,{\bm b})},\nonumber
\end{eqnarray}
where $n_{\rm coll}({\bm x}_T, {\bm b})=\sigma_{\rm inel}T_A({\bm x}_T+{\bm b}/2)T_A({\bm x}_T-{\bm b}/2)$ is the density of binary collisions with impact parameter $\bm b$. $T_{A(B)}$ is the thickness functions~\cite{Miller:2007ri} of nucleus $A(B)$ and $\sigma_{\rm inel}$ is the inelastic scattering cross-section between nucleons. For Pb-Pb collisions at $\sqrt{s_{\rm NN}}=5.02~\rm TeV$, $\sigma_{\rm inel}=68 \rm mb$~\cite{ALICE:2012fjm}. 
$\mathcal{R}$ is the shadowing effect, which is calculated from the EPS09 NLO package~\cite{Eskola:2009uj},  

\begin{figure*}[!htb]
\includegraphics[width=0.85\textwidth]{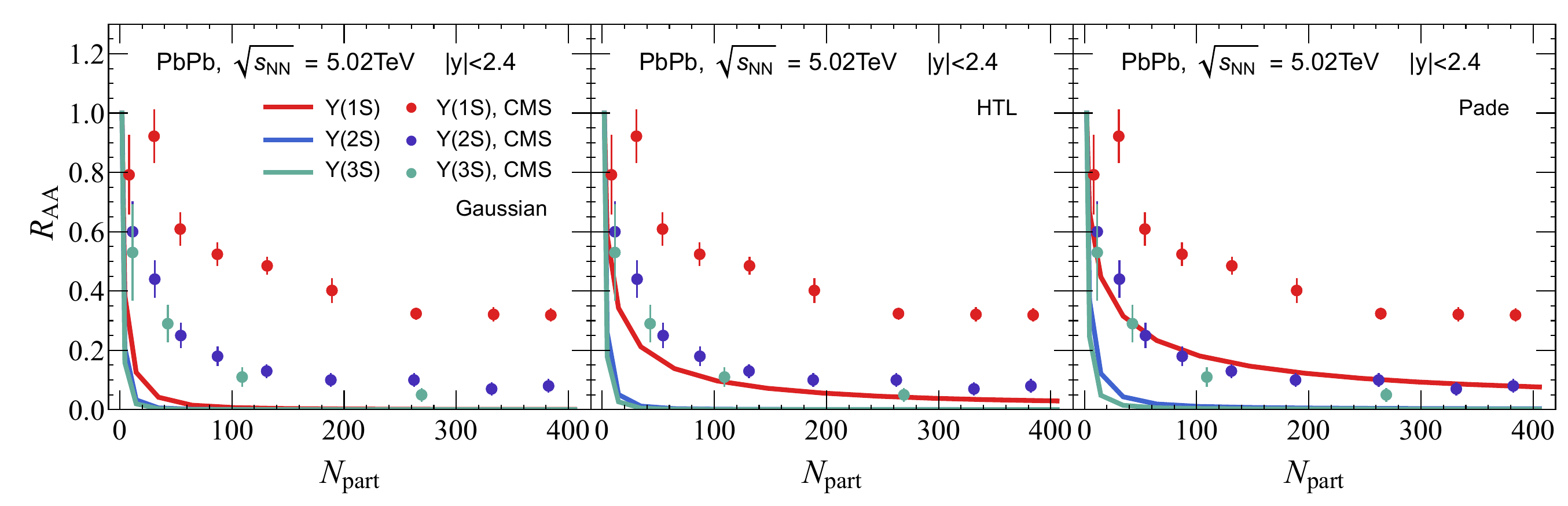}\\
\includegraphics[width=0.85\textwidth]{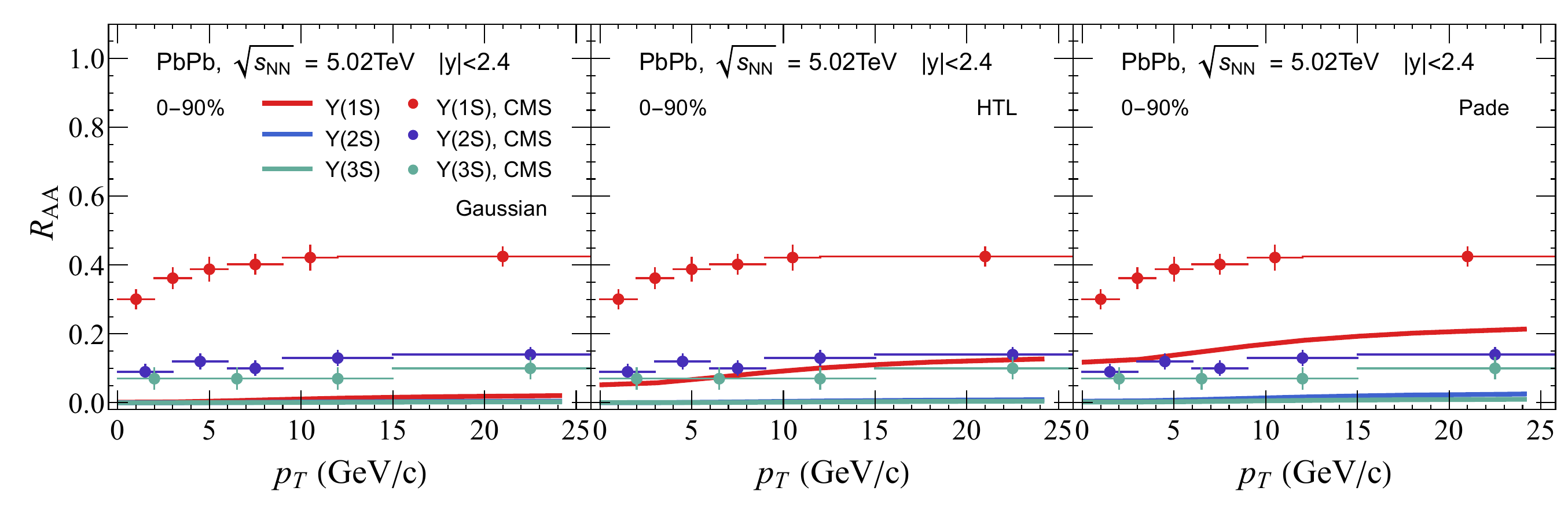}
\caption{The nuclear modification factors of bottomonium $\Upsilon(1s, 2s, 3s)$ as a function of the number of participants $N_{\rm part}$ (upper) and the transverse momentum $p_T$ (lower) in the central rapidity of Pb-Pb collisions at $\sqrt{s_{\rm NN}}=5.02 \rm TeV$. The complex potential is taken as The in-medium potential is taken from Gaussian fit (left), HTL fit (middle), and the Pad\'e fit (right), respectively. The experimental data are cited from CMS~\cite{CMS:2018zza,CMS:2023lfu}.
}
\label{fig-htl-pade}
\end{figure*}
\begin{figure*}[!htb]
\includegraphics[width=0.85\textwidth]{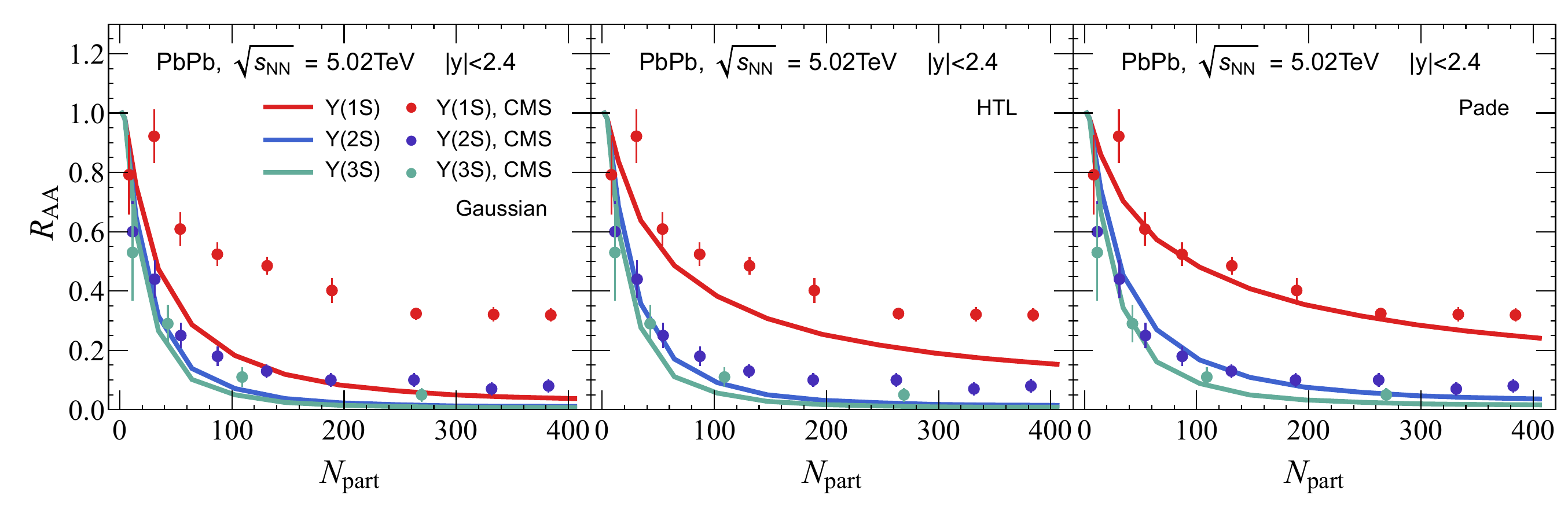}
\includegraphics[width=0.85\textwidth]{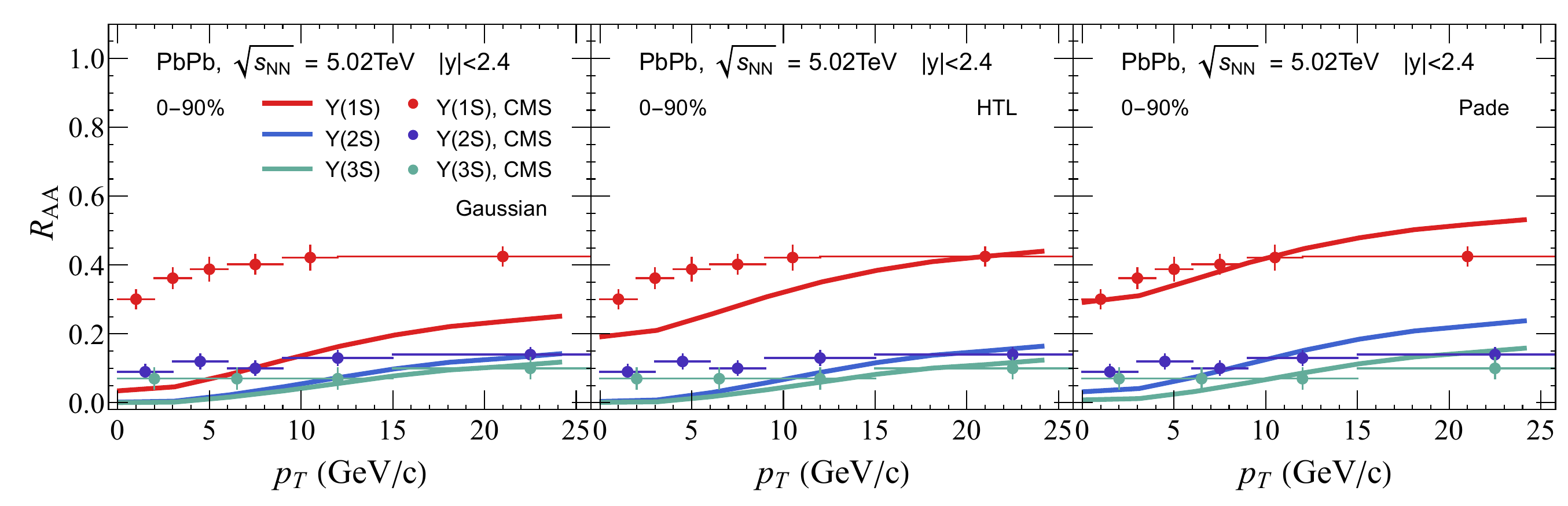}
\caption{Same as Fig.~\ref{fig-htl-pade} but with the formation time of bottomonium $\tau_{\Upsilon}=3 \rm fm/c$.}
\label{fig-htl-pade-tau-3}
\end{figure*}

The number of initial produced bottomonium states will be changed in the hot medium due to either the transition to other states or the dissociation, which describe the transition from the color singlet state to the octet state and is encoded in the imaginary potential. The nuclear modification factor reveals the number change directly. 
Due to the decays from various states, the final nuclear modification factor for $\Upsilon(nS)$ can be expressed as follows,
\begin{eqnarray}
\label{eq-promptRAA}
R_{AA}(\Upsilon(ns))
= {\sum_{nl} \langle |c_{nl}(t)|^2\rangle \sigma_{pp}^{nl}
\mathcal{B}_{nl\rightarrow ns}\over \sum_{nl}
\langle |c_{nl}(t_0)|^2\rangle  \sigma_{pp}^{nl} \mathcal{B}_{nl\rightarrow ns}}.
\end{eqnarray}
where $\langle |c_{nl}(t_0)|^2\rangle$ in the denomenator characterizes the initial production of bottomonium eigenstates without both hot and cold nuclear matter effects, so it can also be calculated with Eq.~\eqref{eq-dRAA} by substituting $\mathcal{R}=1$ into the equation.
The direct production cross section of different bottomonium eigenstates $\sigma_{pp}^{nl}$ is extracted from the experimental data in p-p collisions, $\sigma_{pp}(1s,1p,2s,2p,3s)=(37.97,44.20,18.27,37.68,8.21)$ nb~\cite{Wen:2022yjx,Islam:2020bnp} without feed-down contribution. While the branching ratio of the decay $\mathcal{B}_{nl\rightarrow 1s}$ is taken from Particle Data Group~\cite{ParticleDataGroup:2018ovx}. The decay values from higher to lower states are taken as $\mathcal{B}_{(1p,2s,2p,3s)\rightarrow 1s}=(0.247,0.264,0.088,0.065)$, $\mathcal{B}_{(2p,3s)\rightarrow 2s}=(0.124,0.106)$ and $\mathcal{B}_{(2p,3s)\rightarrow 1p}=(0.0065,0)$.   Similar equation is employed to get the prompt $R_{AA}(2S)$ and $R_{AA}(3S)$.

Now let's see the results obtained by taking the three different in-medium potentials as discussed in the last section.
First, we assume that all bottomonium eigenstates have been formed at $\tau_\Upsilon=\tau_0=0.6\rm fm/c$, which is the start time of the hydrodynamics, and subsequently, their internal evolution is governed by the time-dependent Schr\"odinger equation~\eqref{fun-rad-sch} with complex potentials.
The nuclear modification factors of bottomonium as a function of the number of participants ($N_{part}$) and transverse momentum ($p_T$) are calculated with three different potentials are shown in Fig.~\ref{fig-htl-pade}. 
Due to the very large imaginary potentials in the Gaussian fit, bottomonium suppression is significant and far below the experimental data points. When the imaginary potential is taken as HTL or Pad\'e fit, the bottomonium suppression becomes a bit weaker, but still underestimates the experimental data, see Fig.~\ref{fig-htl-pade}. 
With this formation time, the bottomonium stays the longest time in the hot medium and suffers the largest dissociation in the hot medium.

In another limit, all bottomonium states are formed with a large value of formation time $\tau_\Upsilon=3~ \rm fm/c$. Other setups are exactly the same as before. In this case, the dissociation rates of bottomonium states are zero before $3~ \rm fm/c$ even if they are in the QGP medium. The results are shown in Fig.~\ref{fig-htl-pade-tau-3}. 
The hot medium suppression becomes weaker for all potentials.
We find the calculations with Pad\'e fit can well explain the 
data points. 

In our studies, the real part of the potential does not change the nuclear modification factors of bottomonium evidently. The imaginary part of the potential, which contributes an exponential damping factor in the wave function of $b\bar b$, can significantly affect the final results. In our recent framework, the dissociation from the color singlet state to the octet state is involved in the imaginary potential in the Hamiltonian, while the inverse process from the octet to singlet state has been temporarily neglected, which can be interpreted as regeneration from a correlated $b\bar b$ dipole. Now we want to discuss the importance of transition from octet to singlet state. If one starts to evolve the $b\bar b$ states from a singlet state, then of course the transition from singlet to octet is dominated while the inverse process is largely suppressed as used in previous studies~\cite{Brambilla:2023hkw,Brambilla:2016wgg}. After a long time evolution, the transition will approach detail balance. However, if one starts to evolve the $b\bar b$ states from an octet state, the transition from the octet to the singlet state is definitely important and cannot be neglected.  

In summary, we employ the Schr\"odinger equation with complex potentials obtained recently from lattice QCD to study the static properties and dynamical evolution of bottomonium states in the quark-gluon plasma, created in relativistic heavy ion collisions. 
The mass and thermal width of the bottomonium state reveal the real and imaginary parts of the in-medium potential, respectively. 
The evolution of bottomonium wave functions is studied under different potentials, where the corresponding nuclear modification factors are calculated. Due to the large imaginary potential in Gaussian fit, the nuclear modification factors are reduced to almost zero. While with the other two complex potentials from the HTL fit and the Pad\'e fit, the medium suppression becomes weaker than the case of the Gaussian fit, but still lower than the experimental data. 
We find with a large formation time, the $R_{AA}$ of bottomonium can be well described with the heavy quark potential extracted by the Pad\'e fit, which shows no color screening in the real part potential. This indicates the importance of the bottomonium formation time. This framework helps to establish the correspondence between the in-medium heavy quark potential from first-principle-based lattice QCD and experimental observables about heavy quarkonium, aiding in the understanding of heavy quark physics.

\vspace{1cm}
{\bf Acknowledgement: } This work is supported by the National Natural Science Foundation of China
(NSFC) under Grants No. 12175165. JX is supported by the European Union's Horizon 2020 research and innovation program under grant agreement No. 824093 (STRONG-2020).

\bibliographystyle{apsrev4-1.bst}
\bibliography{Ref.bib}

\end{document}